\documentclass[prl,showpacs,amsmath,twocolumn]{revtex4}
\usepackage{revsymb}
\usepackage{graphicx}
\usepackage{bm}

\begin{document}
\newcommand{\Tr}{\mathop{\mathrm{Tr}}\nolimits}
\renewcommand{\Re}{\mathop{\mathrm{Re}}\nolimits}

\title{Purification through Zeno-like Measurements}
\author{Hiromichi Nakazato}
\email{hiromici@waseda.jp}
\author{Tomoko Takazawa}
\author{Kazuya Yuasa}
\email{yuasa@hep.phys.waseda.ac.jp}
\affiliation{Department of Physics, Waseda University, Tokyo
169--8555, Japan}
\date{January 8, 2003}
\begin{abstract}
A series of frequent measurements on a quantum system (Zeno-like
measurements) is shown to result in the ``purification'' of
another quantum system in interaction with the former.
Even though the measurements are performed on the former system,
their effect drives the latter into a pure state, irrespectively
of its initial (mixed) state, provided certain conditions are
satisfied.
\end{abstract}
\maketitle

It is well known that unstable particles or quantum states
display a peculiar behavior at short and long
times~\cite{NNPijmpb96}.
Phenomenologically, they are known to decay exponentially and
this is well confirmed experimentally~\cite{CKYptp93}.
Short-time deviations were only observed very
recently~\cite{WBFMMNSRnature97}.
The deviations from the familiar exponential decay law are
unavoidable consequences of the quantum dynamics both at short
and long times, and the derivation of the exponential decay law
itself is not a trivial matter in quantum mechanics.
These deviations reflect the unitarity of the time evolution
operator or the time reversal symmetry of the Schr\"odinger
equation at short times and the lower boundedness of the
Hamiltonian or the stability of the vacuum at long times.
See, e.g., Ref.~\cite{NNPijmpb96} for a review.

The quantum behavior of unstable states at short times has been
one of the central issues of investigation and discussion in
recent years, since it is closely connected to the so-called
quantum Zeno effect (QZE)~\cite{QZE,QZEMisraSudarshan}, where the
act of measurement~\cite{WH83} (usually represented by the von
Neumann projection or the generalized spectral
decomposition~\cite{PNpra94}) affects in an essential way the
dynamics of the measured system and results in a hindrance of the
decay process.
The first attempt at the experimental observation of the QZE in
an atomic transition process~\cite{IHBWpra90}, following Cook's
theoretical work~\cite{Cps88}, has triggered heated discussions
on this subject.
Furthermore, another exciting experiment has been reported very
recently: the observation of the QZE (and also of the inverse
QZE~\cite{FNPprl01}) in an atomic tunneling
process~\cite{MGRprl01}, which is the first experimental
observation of the (inverse) QZE in a truly unstable quantum
system, unlike in the previous experiment~\cite{IHBWpra90}
performed on an oscillating system.

In this Letter, we will shed new light on another (and so far not
well explored) feature of the quantum dynamics with measurements,
closely related to the QZE\@.
Notice first that the system under consideration cannot be
considered completely isolated and usually interacts with other
systems.
Therefore, it would be interesting and maybe more realistic to
consider the case where the measurement, represented by a von
Neumann projection for simplicity, is \textit{not performed on
the total system, but only on the system of interest}.
Here, we consider such measurements and address the following
point:
\textit{How does a series of frequent measurements on a system
affect the dynamics of another system in interaction with the
former?}
Under frequent measurements performed only on the former system,
the latter evolves away from its initial state.
We shall show that such measurements can result in a
``purification'' phenomenon.
That is, a series of frequent measurements on system A,
represented by projections on a given (usually pure) state of A,
makes the state of system B, which interacts with A and is
initially in \textit{any} (mixed) state, approach a pure state,
if certain conditions prescribed below are satisfied.

Let a total quantum system A+B be described by a Hamiltonian $H$
of the form
\begin{equation}
H=H_A+H_B+H_\text{int},
\label{eq:totalH}
\end{equation}
where $H_{A(B)}$ stands for a free Hamiltonian of system A(B) and
$H_\text{int}$ for an interaction.
We prepare system A in its initial (pure) state
$|\phi\rangle\langle\phi|$ at time $t=0$.
The initial state of system B, denoted by $\rho_B$, can be
arbitrary.
The initial state of the total system is
\begin{equation}
\rho_0=|\phi\rangle\langle\phi|\otimes\rho_B
\label{eq:rho0}
\end{equation}
and its dynamics is governed by the Hamiltonian~(\ref{eq:totalH})
unless it is interrupted by a series of measurements on system A,
each of which is represented by a projection operator
$\mathcal{O}=|\phi\rangle\langle\phi|\otimes\openone$, performed
at time intervals $\tau$.
Notice that this operator, even if it is a \textit{bona fide}
projection operator (we assume that $|\phi\rangle$ is
normalized), does not return the time-evolved total system to its
initial state.
The projection is partial, in the sense that only the state of
system A is set back to its initial state and that of system B is
not initialized, even though the dynamics of B is certainly
affected.

After $N$ such measurements have been done, the survival
probability of finding system A still in its initial state is
represented by
\begin{eqnarray}
P^{(\tau)}(N)
&=&\Tr\Bigl[
 (\mathcal{O}e^{-iH\tau}\mathcal{O})^N\rho_0
 (\mathcal{O}e^{iH\tau}\mathcal{O})^N
\Bigr]\nonumber\\
&=&\Tr_B\Bigl[
   \bigl(V_\phi(\tau)\bigr)^N
   \rho_B
   \bigl(V_\phi^\dagger(\tau)\bigr)^N
\Bigr].
\label{eq:PNtau}
\end{eqnarray}
Notice that the quantity
$V_\phi(\tau)\equiv\langle\phi|e^{-iH\tau}|\phi\rangle$ is an
operator acting on the Hilbert space of system B\@.
The density operators of the total and B systems read
\begin{subequations}
\label{eq:rhoB}
\begin{eqnarray}
\rho^{(\tau)}(N)
&=&(\mathcal{O}e^{-iH\tau}\mathcal{O})^N\rho_0
   (\mathcal{O}e^{iH\tau}\mathcal{O})^N/P^{(\tau)}(N)\nonumber\\
&=&|\phi\rangle\langle\phi|\otimes\rho_B^{(\tau)}(N),
\end{eqnarray}
\begin{equation}
\rho_B^{(\tau)}(N)
=\bigl(V_\phi(\tau)\bigr)^N\rho_B
 \bigl(V_\phi^\dagger(\tau)\bigr)^N/P^{(\tau)}(N),
\end{equation}
\end{subequations}
respectively.
We only collect the right outcomes of measurements: this is
implicit in the normalization factors in~(\ref{eq:rhoB}).
Experimentally, this means that after each measurement, only
those events will be retained in which system A has been found in
its initial state.

In the ordinary situation, one performs infinitely frequent
measurements by taking $N\to\infty$ and $\tau\to0$, keeping
$N\tau=T$, a finite nontrivial value; one easily checks that the
ordinary QZE~\cite{NNPijmpb96} appears in this case and the
survival probability $P^{(\tau)}(N)$ increases as $N$ becomes
large, approaching unity in the $N\to\infty$
limit~\cite{QZEMisraSudarshan}.
At the same time, the dynamics of system B becomes unitary in
this limit, and this is an example of the so-called ``quantum
Zeno dynamics''~\cite{FKPSpla99}.
However, we stress that our interest lies in a different
situation: we keep the time interval $\tau$ between measurements
finite and nonvanishing.
If $N$ were taken to be $\infty$, the survival probability
$P^{(\tau)}(N)$ would decay out completely for such $\tau\neq0$,
but we are interested in the asymptotic behavior of the state of
system B for large but \textit{finite} values of $N$\@.
We expect that the effect of repeated measurements on system A
would modify the dynamics of system B through its interaction
with the measured system A, even if B has never been directly
measured.
To examine this idea, we need to clarify the asymptotic behavior
of the state of system B, $\rho_B^{(\tau)}(N)$, for large $N$\@.

It is clear that the behavior of $\rho_B^{(\tau)}(N)$ is governed
by the operator $V_\phi(\tau)$ in~(\ref{eq:rhoB}).
Let us consider its eigenvalue problem.
Since this operator is \textit{not hermitian},
$V_\phi^\dagger(\tau)\neq V_\phi(\tau)$, in general, we need to
set up both the right- and left-eigenvalue problems
\begin{equation}
V_\phi(\tau)|u_n)
=\lambda_n|u_n),\qquad
(v_n|V_\phi(\tau)=\lambda_n(v_n|.
\label{eq:uvn}
\end{equation}
The eigenvalue $\lambda_n$ is in general complex-valued.
Let us assume that the spectrum of the operator $V_\phi(\tau)$ is
discrete and nondegenerate, and its eigenvectors form an
orthonormal complete set in the following sense
\begin{equation}
\sum_n|u_n)(v_n|=\openone,\qquad(v_n|u_m)=\delta_{nm}.
\label{eq:uv&vu}
\end{equation}
[It will soon become clear that the assumption of the
nondegenerate spectrum is not essential for the following
discussion except for that of the largest (in magnitude)
eigenvalue $\lambda_\text{max}$.]
The operator itself is expanded in terms of its eigenvectors
\begin{equation}
V_\phi(\tau)=\sum_n\lambda_n|u_n)(v_n|
\label{eq:lambdauv}
\end{equation}
and we obtain
\begin{equation}
\bigl(V_\phi(\tau)\bigr)^N=\sum_n\lambda_n^N|u_n)(v_n|.
\label{eq:lambdaNuv}
\end{equation}
One can show~\cite{NTY02pra?} that the absolute value of the
eigenvalue $\lambda_n$ satisfies the inequality
$0\le|\lambda_n|\le1$, $\forall n$, which reflects the unitarity
of the time-evolution operator.
It is now evident that, in the large $N$ limit, the
operator~(\ref{eq:lambdaNuv}) is dominated by a single term
\begin{equation}
\bigl(V_\phi(\tau)\bigr)^N\xrightarrow{\text{large}\ N}
\lambda_\text{max}^N|u_\text{max})(v_\text{max}|,
\label{eq:Nlimit}
\end{equation}
where $|u_\text{max})$ and $(v_\text{max}|$ are the eigenvectors
belonging to $\lambda_\text{max}$, provided the largest (in
magnitude) eigenvalue $\lambda_\text{max}$ is unique, discrete,
and nondegenerate.

Thus we reach the conclusion, under the assumption of unique,
discrete, and nondegenerate $\lambda_\text{max}$, that, in the
large $N$ limit with a nonvanishing $\tau$, the state of system B
in interaction with system A, on which $N$ measurements are
performed at time intervals $\tau$, asymptotically approaches the
pure state $|u_\text{max})$
\begin{equation}
\rho_B^{(\tau)}(N)
\xrightarrow{\text{large}\ N}
|u_\text{max})(u_\text{max}|/(u_\text{max}|u_\text{max}),
\label{eq:rhouu}
\end{equation}
with probability
\begin{equation}
P^{(\tau)}(N)
\xrightarrow{\text{large}\ N}
|\lambda_\text{max}|^{2N}(u_\text{max}|u_\text{max})
(v_\text{max}|\rho_B|v_\text{max}).
\label{eq:limitP}
\end{equation}
Notice that the final pure state $|u_\text{max})$ is independent
of the choice of the initial state of system B, i.e., any initial
(mixed) state shall be driven to the unique pure state
$|u_\text{max})$ by repeated measurements performed on the other
system A\@.
Since the asymptotic state $|u_\text{max})$ is one of the
eigenstates of the operator $V_\phi(\tau)$, we have the
possibility of adjusting the interaction strength and the
measurement interval and of choosing an appropriate initial state
$|\phi\rangle$ for system A so that a desired pure state
$|u_\text{max})$ is realized in system B after a large number of
measurements on A [as long as the probability $P^{(\tau)}(N)$
does not become meaninglessly small].
This discloses another feature of the Zeno phenomenon: the action
of the quantum Zeno-like measurements dramatically affects the
dynamics of system B of interest.

A few comments are in order.
First, the existence of a unique, discrete, and nondegenerate
$\lambda_\text{max}$ of the operator $V_\phi(\tau)$, which has
been assumed here, is essential for this purification mechanism.
Even though this condition is satisfied for some systems with a
discrete spectrum as will be shown in the examples below, some
definite mathematical criteria for its validity have yet to be
clarified.
In particular, it remains open if the present purification
mechanism can be applied to systems with continuous spectra.
Nevertheless at the same time, it would be worth stressing that
not a few discrete quantum systems, including 2- or 3-level
systems which play important roles in the field of quantum
information and computation, certainly fall into the category of
systems with unique, discrete, and nondegenerate
$\lambda_\text{max}$.
Second, it is easy to see that the approach to the final pure
state $|u_\text{max})$ is governed by the ratio between the
largest and the second largest (in magnitude) eigenvalues of the
operator $V_\phi(\tau)$.
It is possible that as the number of degrees of freedom
increases, the eigenvalues $\lambda_n$ distribute more closely to
each other, which would make the present purification process
less effective.
Lastly, even though the measurements are repeated many times,
like in the \textit{bona fide} Zeno case, the present scheme does
not explicitly rely on the peculiar quadratic behavior of quantum
systems at short times.
As far as the essential assumption on the spectrum of the
operator $V_\phi(\tau)$ is satisfied, there will be no limit on
the time interval $\tau$.
Notice that the repetition of one and the same quantum
measurement is crucial here.

Let us illustrate the above conclusion in a simple but still
nontrivial model.
We consider two single-mode harmonic oscillators $a$ and $b$, in
interaction in the rotating-wave approximation.
The total Hamiltonian reads
\begin{equation}
H=\Omega a^\dagger a+\omega b^\dagger b
+ig(a^\dagger b-ab^\dagger),
\label{eq:Hab}
\end{equation}
where the frequencies $\Omega$ and $\omega$ and the coupling
constant $g$ are real parameters.
The spectrum is discrete.
We prepare system A (oscillator $a$) in some definite pure state
$|\phi\rangle$ (typically a number state $|n_a\rangle$ or a
coherent state $|\alpha\rangle$) at time $t=0$ and let it evolve
under the above Hamiltonian.
Then the initial state of system A starts to evolve towards other
states owing to the coupling to system B (oscillator $b$), the
initial state of which can be arbitrary.
The state of oscillator $a$ is projected onto its initial state
$|\phi\rangle$ at each measurement, and the interval between
measurements $\tau$ is taken small, compared with the typical
time scales of the system, e.g., $2\pi/\delta$ in~(\ref{eq:ABC})
below.

The eigenvalue problem~(\ref{eq:uvn}) of the relevant operator
$V_\phi(\tau)$ is solved exactly in this case.
Indeed, since the time-evolution operator $e^{-iH\tau}$ can be
factorized as
\begin{equation}
e^{-iH\tau}
=e^{Aa^\dagger b}e^{Ba^\dagger a}e^{Cb^\dagger b}
e^{-Aab^\dagger},
\label{eq:decomp}
\end{equation}
in terms of the $\tau$-dependent functions
\begin{subequations}
\label{eq:ABC}
\begin{equation}
A=\frac{(g/\delta)\sin\delta\tau}%
{\cos\delta\tau+i[(\Omega-\omega)/2\delta]\sin\delta\tau},
\end{equation}
\begin{equation}
B=-\frac{i}{2}(\Omega+\omega)\tau-\ln\!\left[
\cos\delta\tau+i\frac{\Omega-\omega}{2\delta}\sin\delta\tau
\right],
\end{equation}
\begin{equation}
C=-\frac{i}{2}(\Omega+\omega)\tau+\ln\!\left[
\cos\delta\tau+i\frac{\Omega-\omega}{2\delta}\sin\delta\tau
\right],
\end{equation}
\end{subequations}
where $\delta=\sqrt{g^2+(\Omega-\omega)^2/4}$, we easily find the
eigenvectors $|u_n)$ and $(v_n|$ of the operator $V_\phi(\tau)$,
once the initial state $|\phi\rangle$ of oscillator $a$ is
specified.

If we prepare oscillator $a$ in the number state $|n_a\rangle$ at
$t=0$, the relevant operator is calculated to be
\begin{eqnarray}
V_{n_a}(\tau)
&=&\sum_{k=0}^{n_a}\frac{n_a!}{[(n_a-k)!]^2k!}
 e^{kB}(-A^2e^C)^{n_a-k}\nonumber\\
& &\phantom{\sum_{k=0}^{n_a}}{}\times e^{Cb^\dagger b}
\prod_{\ell=1}^{n_a-k}
 (b^\dagger b+\ell),
\label{eq:nUn}
\end{eqnarray}
from which we understand that the number states $|n_b)$ of
oscillator $b$ constitute the set of eigenvectors of the
operator~(\ref{eq:nUn}).
Therefore the state of oscillator $b$ is driven to a number state
\textit{irrespectively} of its initial state, when the coupled
oscillator $a$ is repeatedly confirmed to be in the number state
$|n_a\rangle$.
The state of system B is \textit{purified} into a number state.

On the other hand, when oscillator $a$ is prepared in a coherent
state $|\alpha\rangle$ and found to be in this state at every
$\tau$, the relevant operator is rearranged to be
\begin{equation}
V_\alpha(\tau)
=e^{-[1-e^B-A^2/(1-e^{-C})]|\alpha|^2}e^{D(b^\dagger,b)},
\label{aUa}
\end{equation}
where the operator $D(b^\dagger,b)$ is expressed as
\begin{equation}
D(b^\dagger,b)
=C\left[b^\dagger+\frac{A\alpha^*}{1-e^{-C}}\right]
  \left[b-\frac{A\alpha}{1-e^{-C}}\right].
\label{eq:opD}
\end{equation}
It is easily understood that the state of oscillator $b$
approaches a coherent state $b|\beta)=\beta|\beta)$ with
$\beta=A\alpha/(1-e^{-C})$, since this is the right-eigenvector
of $D(b^\dagger,b)$ belonging to zero eigenvalue and therefore
that of $V_\alpha(\tau)$ belonging to the largest (in magnitude)
eigenvalue~\cite{Eigenvalues}.
The state of system B is again \textit{purified} into (another)
pure state $|\beta)$.

\begin{figure}[t]
\includegraphics[width=0.42\textwidth]{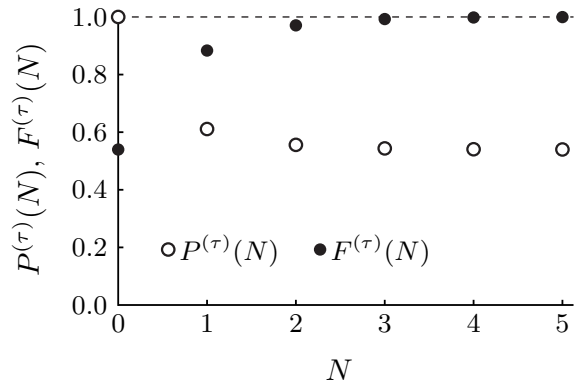}
\caption{Probability $P^{(\tau)}(N)$ and fidelity $F^{(\tau)}(N)$
for the model~(\ref{eq:Hab}) when the initial state of system A,
onto which $N$ measurements are performed, is a coherent state
$|\alpha\rangle$ and that of system B is $\rho_B\propto
e^{-\omega b^\dag b/k_BT}$ (thermal, i.e., maximally mixed).
The parameters are taken to be $\omega=1$, $g=0.2$, $T=1$,
$\alpha=0.5$ and $\tau=2\pi/[(\Omega+\omega)/2+\delta]\simeq5.24$
in units such that $\hbar=k_B=\Omega=1$.
$\tau$ is tuned so as to satisfy the condition
$|\lambda_\text{max}|=1$~\cite{NTY02pra?}, and the ratio of the
second largest (in magnitude) eigenvalue to the largest one
$\lambda_\text{max}$ is $|e^C|=\sqrt{1-(g/\delta)^2
\sin^2\!\delta\tau}\simeq0.37$~\cite{Eigenvalues}.}
\label{fig:ProbabilityFidelity}
\end{figure}
In Fig.~\ref{fig:ProbabilityFidelity}, the survival probability
$P^{(\tau)}(N)$ and the so-called fidelity
\begin{equation}
F^{(\tau)}(N)
=(u_\text{max}|\rho_B^{(\tau)}(N)|u_\text{max})/%
(u_\text{max}|u_\text{max})
\end{equation}
are shown as functions of the number of measurements $N$ for the
case (\ref{aUa})--(\ref{eq:opD}) with a particular choice of
parameters.
In order to make the purification procedure more effective, it is
preferable that (i)~the magnitude of the largest eigenvalue of
the operator $V_\phi(\tau)$ be close to one,
$|\lambda_\text{max}|\simeq1$, which maintains the probability
$P^{(\tau)}(N)$ large enough even for large $N$
[see~(\ref{eq:limitP})], and (ii)~the other eigenvalues be all
small (in magnitude) compared with $|\lambda_\text{max}|$, in
order to realize a faster approach to the final pure state
$|u_\text{max})$.
For this purpose, one may adjust the relevant parameters, such as
the interval between measurements $\tau$, the strength of the
interaction $g$, and the state $|\alpha\rangle$ onto which system
A is projected.
The condition $|\lambda_\text{max}|\simeq1$ is satisfied in
general if the interval $\tau$ is taken to be small enough as in
the ordinary Zeno measurements, but one can optimize this
procedure and find better values of $\tau$, not necessarily very
small~\cite{NTY02pra?}, that satisfy both conditions~(i)
and~(ii).
See Fig.~\ref{fig:ProbabilityFidelity} and its caption, where
$\tau$ is tuned so as to satisfy the conditions~(i) and~(ii) for
the case (\ref{aUa})--(\ref{eq:opD}), and the purification
mechanism becomes very effective after only $N=2$ steps.

The above arguments clearly and explicitly show how the action of
repeated measurements (projections) on one system A can affect
the dynamics of the other system B in interaction with A\@.
Interestingly enough, even though the effect of the measurement
on the latter system B is indirect, its influence is far-reaching
if the measurement is repeated many times: irrespectively of its
initial (mixed) state, the state of system B is \textit{purified}
towards a pure state, provided the conditions on the spectrum of
$V_{\phi}(\tau)$ are satisfied.
The final state of system B is prescribed by the total
Hamiltonian, the pure state (usually taken to be the initial
state) onto which system A is projected by the measurement, and
the time interval between successive measurements.
This opens a new possibility on how to control the state of a
quantum system on which we have no direct access.
If another system under our control can be coupled to the former
system, we would only have to decide which state has to be
measured on the controllable system.
After such measurements are performed many times at the
prescribed time intervals, the desired pure state would be
realized with some probability in the system beyond our control.

Purification of quantum states is now considered to be one of the
key technologies for quantum information and computation, and is
being widely explored~\cite{QuantumInfo} (especially in the
context of ``entanglement purification'').
Compared to some other procedures, the idea here is rather
simple: one has only to repeat the same measurements.
The objects to which the present method is applicable are general
and not restricted to ``qubits.''
Furthermore, it is worth emphasizing the versatility of this
procedure, i.e., the possibility to adjust the target pure state,
the balance between fidelity and probability yield, and so on.
These issues would deserve further study, for example, in the
context of quantum information and computation.

\begin{acknowledgments}
The authors acknowledge useful and helpful discussions with
Prof.~Ohba, Prof.~Accardi, and Dr.~Imafuku.
This work is partly supported by a Grant-in-Aid for Priority
Areas Research (B) from the Ministry of Education, Culture,
Sports, Science and Technology, Japan (No.~13135221), and by a
Waseda University Grant for Special Research Projects
(No.~2002A-567).
\end{acknowledgments}



\begin{thebibliography}{99}
\bibitem{Eigenvalues}
The eigenvalues of the operator $V_\alpha(\tau)$ in
Eq.~(\ref{aUa}) are given by
$\lambda_n=e^{-[1-e^B-A^2/(1-e^{-C})]|\alpha|^2}e^{nC}$
($n=0,1,2,\ldots$), and $\lambda_\text{max}=\lambda_0$ as long as
$\Re C\neq0$.
\end{thebibliography}
\end{document}